\documentclass[11pt]{article}
\pdfoutput=1
\makeatletter
\def\ps@headings{%
\def\@oddhead{\mbox{}\scriptsize\rightmark \hfil \thepage}%
\def\@evenhead{\scriptsize\thepage \hfil \leftmark\mbox{}}%
\def\@oddfoot{}%
\def\@evenfoot{}}
\makeatother
\usepackage{latexsym}
\usepackage[latin1]{inputenc}
\usepackage{fullpage}
\usepackage{epsfig}
\usepackage{times}
\usepackage{amsmath,amssymb,amsfonts,amsthm}
\usepackage{xr-hyper}
\usepackage{url}
\usepackage{newlfont}
\usepackage[usenames]{color}
\usepackage{color}
\usepackage{graphics}                 
\usepackage{graphicx}
\usepackage{tikz}
\usepackage{rotating} 
\usepackage{epsfig}
\usepackage{epstopdf}
\usepackage{verbatim} 
\usepackage{algorithm}
\usepackage{algorithmic}
\usepackage{xspace}

\makeatletter

\newcommand{\LyX}{L\kern-.1667em\lower.25em\hbox{Y}\kern-.125emX\spacefactor1000}

\makeatother
\newcommand{\PITc}{\ensuremath{\bar{c}}\xspace}
\newcommand{\PITC}{\ensuremath{\bar{C}}\xspace}
\newcommand{\PITu}{\ensuremath{\bar{u}}\xspace}
\newcommand{\PITU}{\ensuremath{\bar{U}}\xspace}
\newcommand{\PITw}{\ensuremath{\bar{w}}\xspace}
\newcommand{\PITW}{\ensuremath{\bar{W}}\xspace}

\newlength{\defbaselineskip}
\newcommand{\setlinespacing}[1]%
           {\setlength{\baselineskip}{#1 \defbaselineskip}}

\theoremstyle{plain}
\newtheorem{thm}{Theorem}[section]
\newtheorem{cor}[thm]{Corollary}
\newtheorem{lem}[thm]{Lemma}
\newtheorem{prop}[thm]{Proposition}
\theoremstyle{definition}
\newtheorem{defn}{Definition}[section]
\theoremstyle{remark}

\numberwithin{equation}{section}
\renewcommand{\theequation}{\thesection.\arabic{equation}}

\newcommand{\argmin}[1]{{\hbox{$\underset{#1}{\mbox{\emph{argmin}}}\;$}}}

\newcounter{algline}

\newcommand{\nl}{\\ \thealgline \stepcounter{algline}}

\newenvironment{alg}[1]{\setcounter{algline}{1}\begin{tabbing} 123\=678\=123\=678\=123\=678\=123\=678\=123\=678\=  \kill {\bf #1:} \nl}{\end{tabbing}}

\newenvironment{boxfig}[1]{\begin{figure*}[htb]\fbox{\begin{minipage}{0.98\linewidth}
                        \vspace{1em}
                        \makebox[0.025\linewidth]{}
                        \begin{minipage}{0.95\linewidth}
                        #1
                        \end{minipage}
                        \end{minipage}}}{\end{figure*}}

\begin{document}

\graphicspath{{./figures/}}
%
\title{Quid Pro Quo: A Mechanism for Fair Collaboration in Networked Systems\thanks{This research was supported in part by the Comunidad de Madrid grant S-2009/TIC-1692, the Spanish MICINN grant TEC2011-29688-C02, and the National Natural Science Foundation of China grant 61020106002.}}

\author{Agustín Santos\\
Institute IMDEA Networks, Madrid, Spain\\
E-mail: agustin.santos@imdea.org
\and
Antonio Fernández Anta\\
Institute IMDEA Networks, Madrid, Spain\\
E-mail: antonio.fernandez@imdea.org
\and
Luis López Fernández\\
LADyR, GSyC\\
Universidad Rey Juan Carlos, Madrid, Spain\\
E-mail: llopez@gsyc.es }

\date{}

\maketitle

\begin{@twocolumnfalse} \begin{abstract}


Collaboration may be understood as the execution of coordinated tasks (in the most general sense) by groups of users, who cooperate for achieving a common goal. Collaboration is a fundamental assumption and requirement for the correct operation of many communication systems.
The main challenge when creating collaborative systems in a decentralized manner is dealing with the fact that users may behave in selfish ways, trying to obtain the benefits of the tasks but without participating in their execution. 
In this context, Game Theory has been instrumental to model collaborative systems and the task allocation problem, and to design mechanisms for optimal allocation of tasks.
In this paper, we revise the classical assumptions and propose a new approach to this problem. First, we establish a system model based on heterogenous nodes (users, players), and propose a basic distributed mechanism so that, when a new task appears, it is assigned to the most suitable node. The classical technique for compensating a node that executes a task is the use of payments (which in most networks are hard or impossible to implement). Instead, we
propose a distributed mechanism for the optimal allocation of tasks without payments. We prove this mechanism to be robust event in the presence of independent selfish or rationally limited players. Additionally, our model is based on very weak assumptions, which makes the proposed mechanisms susceptible to be implemented in networked systems (e.g., the Internet). 

\end{abstract}\end{@twocolumnfalse}
\setlength{\columnsep}{0.75cm}

\section{Introduction}

Selfish behavior is becoming a subject of great concern and practical importance to network designers \cite{han2012}. 
Game Theory is the approach of preference to face the design of communication systems with (potentially)
selfish entities. This has lead to the proposal of a number of interesting protocols and mechanisms 
for networks based on Game Theory concepts \cite{Bell2000533,citeulike:2624188}.
However, in the study of networks under conventional models, a collection of simplifying assumptions are typically made.
For instance, it is assumed that selfish users are rational, that they are homogeneous,
that they can compute a Nash equilibrium, that their utility function is known, etc.
However, there are many systems in which these assumptions
assumptions are not very realistic. 

In this paper we revisit the study of communication systems with selfish
users (or players), reevaluating and relaxing the above-mentioned common assumptions.
In particular, we propose the problem of analyzing and designing of a fair collaborative
system under a very weak set of game theoretic assumptions.
In this general context, we propose mechanisms to be used to implement this
collaborative system with provable properties, like the fairness of the system and
the truthfulness of its users.
The mechanisms proposed can be applied to such varied technologies as
social and crowd computing, Web 2.0, P2P, opportunistic networks, and cloud technologies.

As mentioned, we abstract the problem to be solved as the fair execution of tasks in a decentralized collaborative system.
The main challenge when creating collaborative systems in a decentralized manner is dealing 
with the fact that system nodes may behave in selfish ways, trying to obtain the 
benefits of the tasks but without participating in their execution. 
(This is the realm of Game Theory, which has been instrumental to model collaborative systems 
and the task allocation problem, and to design mechanisms for optimal allocation of tasks.)
We assume that all nodes have an interest on having the tasks done. However, establishing fair mechanisms for sharing the generated work-load is not immediate. (E.g., in current P2P systems, usually a low fraction of peers assume most of the required effort, and this causes reduced performance, lack of reliability, low incentive to participate for fair users, etc.)
It would therefore be desirable that each node could take the responsibility of the execution of a balanced fraction of the tasks. 

The objective is to establish some kind of protocol to share the task execution costs. For this, we need to consider the concept of ability or \emph{opportunity of execution.} Let us assume that each node has some capacity for timely execution of a given task. This capacity may vary over time and with the type of task. For example, at a given time, a node may have free bandwidth but have full utilization of its CPU, while its situation could be the oposite at another time. Hence, at a particular moment, a node may have greater ability to perform tasks involving communication, while at a later time its situation may change to prefer tasks more intensive in CPU computation.

This opportunity or ability is related with the notion of task execution cost. In other words, we define the cost as some kind of metric measuring the capability of executing a particular task at a given time. Hence, the cost varies from one task to another (even when the task is the same, but at a later time). In Game Theory, closely related to cost, there is the notion of utility. We define the utility as the cost savings associated with a work not done. Hence, given that all nodes are interested in the execution of the tasks, a node gets more utility whenever it avoids running tasks by letting other nodes to do it.

Clearly, when trying to formalize a model based on these notions, a number of problems arise. First, node's costs are only known by the node itself. For external entities it would be difficult to audit or check if a given particular node has more or less CPU capacity. In Game Theory, this concept is called \emph {private information}. For obtaining the private information of a node, the basic mechanism is to directly ask for it and expect the node to declare its value correctly.

For us, each node is a computing node that belongs to a user who can alter her node's behavior for her own benefit (i.e., may declare false costs trying to avoid the execution of tasks). Whenever this happens, we claim that the user acts in a selfish way. This selfishness is one of the factors that may distort the internal workings of a distributed application. The loss of system performance produced by selfish nodes is a parameter to consider and it is called price of anarchy \cite {KoutsoupiasPapadimitriouWorstcaseEquilibria, RoughgardenSelfishRoutingPriceAnarchy}.

Therefore, the problem we face consists of designing a system capable of assigning tasks to nodes so that all the tasks are executed, and the total cost incurred is minimal. When the behavior of nodes is guaranteed to be fair, this is just a simple optimization exercise. However, when nodes may choose whether to be selfish the problem becomes much more complex.
In this paper we propose an algorithm that, basing on game theory principles, solves this problem. We have called this algorithm \textit{Quid Pro Quo Mechanism} (QPQ). The name comes from a Latin expression commonly used by lawyers and which may be translated as \textit {``This for that''} or \textit{``A thing for another''}. This expression is often used when someone makes a job or and waits for and equivalent compensation in exchange. We used this expression since it reflects the spirit of the algorithm: due to the lack of payments in our model, the nodes work for others with the hope that others will work for them in the future.

\subsection{State of the Art}

As described above, the problem addressed in this paper is the allocation of task executions to potentially selfish users. This problem has been extensively studied in the literature. One important related work was carried out by Rosenschein et al. \cite{Rosenschein}, where they define a ``Task Oriented Domain''. Even though they obtain fairly relevant conclusions, they do not shed any light on the specific problem considered here, since their model makes strong assumptions, such as knowledge of the task costs or a bargaining power over time.
Recently, the use of game theory to model selfish behavior in the design of distributed systems has been proposed. Some works have appeared using mechanism design, a branch of mathematics derived form game theory, which provides the required background for the study and design of distributed systems under the action of selfish nodes (see, e.g., \cite {Papadimitriou, jackson03mechanism, JacksonCrashCourse}). 

In this direction, our QPQ algorithm is similar to the mechanism proposed by Jackson et al.~\cite{LinkingCollectiveDecisionsEfficiency}. In that work, they present a new interesting type of mechanism (called \emph {linking mechanism''}) which, instead of offering incentives or payments to players, limits the spectrum of players' responses to a probability distribution known by the game designer. In that paper the authors proved that a linking mechanism is valid when the players' possible decisions are distributed following discrete probabilities. Additionally, the authors show that a linking mechanism can also be used for repeated games. Even though the work of Jackson et al.~is very relevant to the problem we consider, it does not offer a method for the construction of mechanisms when the game is based on unknown continuous probability distributions, as assumed here.
A second work that explores the idea of linking mechanism due to Ferenc \cite{LinkingDecisionsWithMoments}. In that paper, he proposes a mechanism which limits player responses by restricting the first two moments (mean and variance) of the probability distribution, being that distribution known to the designer.
%
%
Both works reflect the main idea behind the concept of linking mechanism: when a game consists of multiple instances of the same basic decision problem (e.g., saying yes or no, choosing among a number of discrete options), it is possible to define selfishness-resistant algorithms by restricting the players' responses to a given distribution. Hence, in that case, the frequency with which a player declares a particular decision is known beforehand.

In the specific areas of computing and communications, it is important to remark that most mechanisms proposed for dealing with selfish agents make unrealistic assumptions\cite{amir-expected}. In this direction, Bauer et al.~\cite{bauer06-assessing} criticize many of these hypotheses, reviewing well-known works \cite{AdhocVCG,feigenbaum02bgpbased,feigenbaum02distributed} to show that they are not applicable in real environments. Specifically, they identify two common strong artificial assumptions:
\begin{enumerate}
\item The assumption that the designer of the algorithms has some knowledge
 about the preferences of the nodes. 
\item The assumption that the interaction among
 players is limited to a single round (while it is well known in the literature that a solution for a
 single round does not necessarily apply when the game is repeated).
\end{enumerate}

\subsection{Contributions}
In this paper, we face the problem of task allocation relaxing these (and other) common hypotheses, so that the obtained results can be applied in real environments. Hence, the contributions of this paper are twofold. First, to the best of our knowledge, this is the first work proposing a linking mechanism solution without prior knowledge of the distribution of the players' decisions, and without a payment system among them. Second, we generalize and improve previous works in the area to provide algorithms which are susceptible of being applied in the context of repeated task execution allocation in real communication and computing systems, even in the presence of selfish or non-rational users.


As we previously claimed, we do not want to restrict our mechanism to a set of unrealistic hypotheses. Instead, we establish a number of requirements that our model must satisfy. These requirements should provide the appropriate flexibility to guarantee the applicability of our results in real environments.

\paragraph{Abstract utility metrics}
We assume, as an abstract notion, that the cost of a executing task to a node\footnote{We will user the terms user, player, and node indistinctly in the rest of the paper.} 
depends on its interest on the task, its opportunity or ability to execute it, or its degree of willingness to cooperate. We need to accept that each node may measure this parameter in its very own metric and units. Hence, for example, a node may decide on the cost of a task according to the occupation of its CPU, but another one may prefer to make it depend on its available bandwidth. In a real scenario, the number of factors that can influence the execution cost of a task can be extremely large. 
In this direction, out model must enable each node to define, in a flexible way, how costs (and utilities) are measured.

\paragraph{No payment system}
Payments are, in its most basic interpretation, a way of exchanging costs. Many existing mechanisms base their incentive schemes on the existence of payments. For payments to be possible, it is necessary that all players manage a common currency reference (euro, dollar, etc.). However, given our previous requirement, it is not clear how we can find that shared currency reference in our model. If a node measures its costs in terms of, for example, reputation, it can hardly ``pay'' to another node that measures its costs on CPU units. Hence, in our work, we assume that payments are not possible.

\paragraph{Player's rationality}
In game theory, most of the existing algorithms require players to be perfectly rational. This means that a player, using the available information, should always be capable of selecting the best strategy (the one that maximizes her utility). However, this is a controversial hypothesis which is suffering much criticism. Accepting this assumption means that players are capable of mathematically calculating all alternatives, which in some cases requires solving complex (NP-hard) problems. Clearly, this is not always feasible for all players. Hence, we commit ourselves to proposing mechanisms suitable for finding quasi-optimal task allocation, even in the presence of rationally limited players.

\paragraph{Incentive to participate}
In relation to players rationality, even in the case in which we are able to find global quasi-optimal task allocation, it is possible that the behavior of rationally limited users may harm the benefit of other players. In this direction, we add a stronger requirement. We force to ensure an incentive to participate in the game to all nodes, independently on whether they are rational or not. 

\paragraph{No central entity}
A final requirements we impose is the capability of the system to work without the existence of any kind of central entity. This means that the proposed mechanisms must be susceptible of being implemented following completely distributed schemes.


\subsection{Structure}

The rest of the paper is structured as follows. In Section~\ref{s:definitions} we provide a formal definition of the problem and define basic terminology. In Section~\ref{s:basic} we present a basic linking mechanism, and evaluate the issues that need to be faced to make it suitable for our problem. In Section~\ref{s:qpq} we present the QPQ mechanism,
and formally prove its properties. In Section~\ref{PracticalQuidProQuo} we describe how QPQ could be used in real environments. Finally, Section~\ref{s:conclusions} concludes the paper.

\section{Definitions}
\label{s:definitions}

To establish a formal framework for the problem, let us provide some definitions.

\begin{defn}[Problem]. The problem of the assignment of tasks is a tuple $\langle T, N, C \rangle$ where:
\begin{enumerate}
\item  $T  = \{t_1, t_2,\ldots\} $ is the (not necessarily finite) set of tasks that are issued to the system over time ($t_k$ is the task issued at time step $k$). We assume tasks to be atomic, independent, and of fixed duration $\sigma$. (For simplicity we will assume $\sigma=1$, i.e, each task takes one time step to be executed.) Note that we assume that complex tasks may be divided into atomic tasks.
\item  $N = \{1,2,\ldots, n\}$ is an ordered list of nodes or players, where $N$ is assumed to be finite,
\item  $(C(t))_{i \in N}$ is a vector of costs (or utilities) where $C_i(t)$ is the cost of executing task $t \in T$ by node $i$. This information is private (only known by node $i$).
\end{enumerate}
\end{defn}

It is important to remark some aspects of the above definitions. First, we assume that the set of tasks is not known beforehand. Tasks appear one by one in a sequence of time steps, which command our discrete time evolution. Hence, the arrival of a new task dictates the start of new a round of our repeated game. We assume that tasks are independent among them and that the execution of a task does not influence the cost of the subsequent ones. Moreover, we force that one task must be completely executed
by the time the next task is issued. For simplicity, we assume that the mechanisms to coordinate the allocation of the tasks take negligible time (with respect to the time step).
Finally, we assume that every node that is assigned a task by the allocation mechanism actually executes the task.

Hence, as tasks are issued, each node $i \in N $ estimates a sequence of costs $C_i (t_1), C_i (t_2), \dots, C_i (t_k), \dots $, which we assume as independent samples of a probability distribution $\sigma_i \in \Delta (S_i)$ characterizing node $i$'s behavior.  In this context, we denote $S_i$ as the distribution support (i.e., the range of values for which the probability is different to zero) and $\Delta (S_i)$ as the set of all possible probability distributions over $S_i$. 
From now on, we will consider that $C_i$ is a real-valued random variable with probability distribution $\sigma_i \in \Delta (S_i)$. To simplify the notation, we define realizations of this random variable as $c_i(t) = C_i (t)$, $t \in T$. When clear from the context, we may remove the task $t$ from the notation  $c_i(t)$, as $c_i$.

Given that all players enjoy the result of any task executed in the system, we can define the utility of a player as the savings obtained by not executing some tasks (i.e. the benefit obtained from participating in the cooperative computing scheme and not making all the work by itself). That is, the utility $u_i (t)$ of node $i$ corresponding to a given task $t$ is given by
\begin{equation}\label{eq:utillitydefinition}
u_i(t) = \begin{cases}
0& \text{if node $i$ executes the task},\\
c_i(t)& \text{otherwise},
\end{cases}
\end{equation}
and the total utility of node $i$ is $u_i=\sum_ {t \in T} u_i (t) $.\footnote {In game theory it is common to add a discount factor ($\delta$) in time. We have assumed it to be equal to $ \delta = 1$.}

We define $U_i$ as the random variable associated to the total utility of node $i$. 
In a similar way, we denote by $W_i$ the real-value random variable associated to the actual player $i$'s executed cost and by $w_i(t)$ its concrete realization for task $t$. Note that each task could be executed or not by a particular player. Hence,
\begin{equation}
\label{eq:c=u+v}
\mathit{E}[C_i] = \mathit{E} [U_i + W_i] = \mathit{E} [U_i] +\mathit{E} [W_i].
\end{equation} 

Finally, we assume that communication between players is \emph{reliable} and \emph{concurrent}. In particular, in the mechanisms we propose all players exchange their values $c_i(t)$. 
We assume that these values are correctly received by the players in a time that is negligible with respect to the time step (hence the reliability property). Additionally, we assume that each player sends its value before receiving the value of any of the other players (hence the concurrence property).

\section{Basic Linking Mechanism}
\label{s:basic}


As mentioned above, a linking mechanism is applicable to repeated games where the decision (also know as message) of players is restricted to a particular known set.
In our problem, the decision is the cost $c_i(t)$ of the task.
%
%
With this concept in mind, let us define our first algorithmic attempt to solve the problem by applying a linking mechanism, presented in Algorithm \ref{alg:SimpleLinkingMechanism}. 

\algsetup{indent=2em}
\begin{algorithm}[h!]
\caption{Simple linking mechanism (code for node $i$, and a generic task $t$, omitted)} \label{alg:SimpleLinkingMechanism}
\begin{algorithmic}[1]
\STATE Estimate and publish the cost  $c_i$ of the task
\STATE Wait to receive the costs $c_j$ from the other players
\FORALL{$j \in N$ } 
	\IF{  not $\mathit{Accepted}(c_j,\mathit{Historic}_j)$}
\STATE $c_j \leftarrow \mathit{Random}(c_{-j})$
\ENDIF
\STATE $\mathit{Historic}_j \leftarrow  \mathit{Historic}_j \cup \{c_j\}$
\ENDFOR
\STATE $d \leftarrow \argmin{j \in N}{c_j}$
 \IF{ $d = i$ } 
 \STATE execute the task
  \ELSE 
  \STATE do nothing (node $d$ will execute the task)
  \ENDIF
\end{algorithmic}
\end{algorithm}


As it can be observed, for each task, each player estimates the cost of computing the task and publishes it. Publication means
broadcasting a message with the cost to all players (although any other means of distribution, like shared memory, can be used).
By assumption, a player sends it costs before it receives any of the others (concurrency, which implies that they do not depend from each other), and all of the costs are correctly received at each player (reliability). 
Then, the algorithm assigns the task to the player that publishes the lowest cost. If players publish their real costs, this will produce that the total utility is maximized. However, this kind of approach could drive selfish users to publish fake costs in order to avoid executing tasks. For this reason, we add an acceptation test. When a published cost is not considered acceptable, then the system generates a random value for the cost of that node on the round. The implementation of this acceptation test will be discussed later, however it is important to remark that it contains the linking part of the mechanism (it depends on the historical values published by that particular node). Just as an example, we can imagine that if we mandate that nodes must publish costs between $0$ and $1$ following a uniform distribution, then we could consider unacceptable values deviating from that distribution. It is also important to note that all nodes use the same acceptance test with the same history. Then, they all accept or reject. Then, if players reject a value $c_j$, the value $\mathit{Random}(c_{-j})$ generated is in fact a value deterministically generated from the set of values $c_{-j} = \cup_{k \neq j} \{c_k\}$, so that all players re-generate the same value for $j$.

Algorithm \ref{alg:SimpleLinkingMechanism} has the objective of providing intuition on how we build our mechanism, but it clearly has several issues that contradict our previously stated requirements. In particular, fair allocation is not guaranteed. For instance, there is not a way of defining a notion of fairness within this algorithm, given that costs may have different meanings for different players. Additionally, given that costs are abstract notions, we cannot have any a-priori information on the shape of their corresponding distributions. So, it is not clear how to implement the acceptance test.

Digging into these problems, it is easy to understand that one of their causes is the fact that, given our requirements, each player has the right of measuring her costs on her preferred metric. (Hence, each player may have different distributions with different supports.) For this reason, cost comparisons cannot be easily made. 
%
Additionally, there is a second aspect that must be addressed. In the literature about linking mechanisms, authors assume that instances of the game (rounds) are simultaneous in time. In this case, defining the acceptance function over the set of values is easier. However, in our case, tasks are issued, and hence players generate their costs, over time. Then,  from the point of view of the designer, it is not clear how to determine the acceptance of a value by comparing with a certain probability distribution. The issue is even worse given the fact that this distribution is not known by the designer. To solve all these problems we propose a novel solution based on applying a transformation over the utility function.

\paragraph{Utility normalization}
Given that the utility is defined as the work not done by a node, we may use as utility function of a node its probability distribution of costs. Once this is done, we may modify Algorithm \ref{alg:SimpleLinkingMechanism} and normalize players' utilities so that they may be compared among each other. To normalize we use a transformation called \emph{Probability Integral Transformation} (PIT). Our idea is to use the known fact that any cumulative probability distribution function has in itself a uniform distribution \cite{AngusProbabilityIntegralTransform}.
More formally, the PIT is defined as
\begin{defn}[Probability Integral Transformation] \label{DEFPIT}
Let $X$ be a continuous random variable with a Cumulative Distribution Function (CDF) $F$; that is $X \sim F$. Then, the probability integral transformation defines a new random variable $Y$ as: $ Y = F(X)$.
\end{defn}
As mentioned above, our interest in the PIT is due to the following lemma.
\begin{lem}[PIT follows uniform distribution]\label{PIT}
Let $X$ be a continuous random variable with CDF $F$, then $F$ follows a uniform distribution on interval $[0,1]$.  That is, the random variable $Y$ defined by the probability integral transformation  $Y = F(X)$ is a normalized uniform distribution.
\end{lem}

Note that  $X$ does not need to be a continuous random variable. In the case that the player's costs follow a discrete distribution, it is still possible to perform a similar transformation called \emph{Generalized Distributional Transform} \cite {MathematicalStatisticsFerguson}, whose properties are equivalent to those of the PIT.

\begin{defn}[Generalized Distributional Transform] \label{DEFGDT}
Let $X$ be a random variable (not necessarily continuous) with a cumulative distribution probability $F$ and let $V \sim U(0,1)$ be a random variable with uniform distribution in $[0,1]$ independent of $X$. The \emph{modified distribution function}  $F (x, \lambda)$ is defined as 
$$F (x, \lambda) = \mathit{Pr} (X <x) + \lambda \mathit{Pr} (X = x).$$  
From this, we can define the general distributional transform of $X$ as $Y = F (X, V)$, which can be proved to be a uniform distribution on the unit interval.
\end{defn}
Proofs of these properties can be found in 
\cite {MathematicalStatisticsFerguson}. Many studies in economics use this definition and its properties, such \cite{Burgert06onthe} or \cite{Ludger20093921}. In our case, to simplify the notation, we just call PIT to both transformations independently on whether the base distribution is continuous or discrete. 

\begin{figure}[t]
  \begin{center}
  \resizebox*{!}{10cm}{\includegraphics{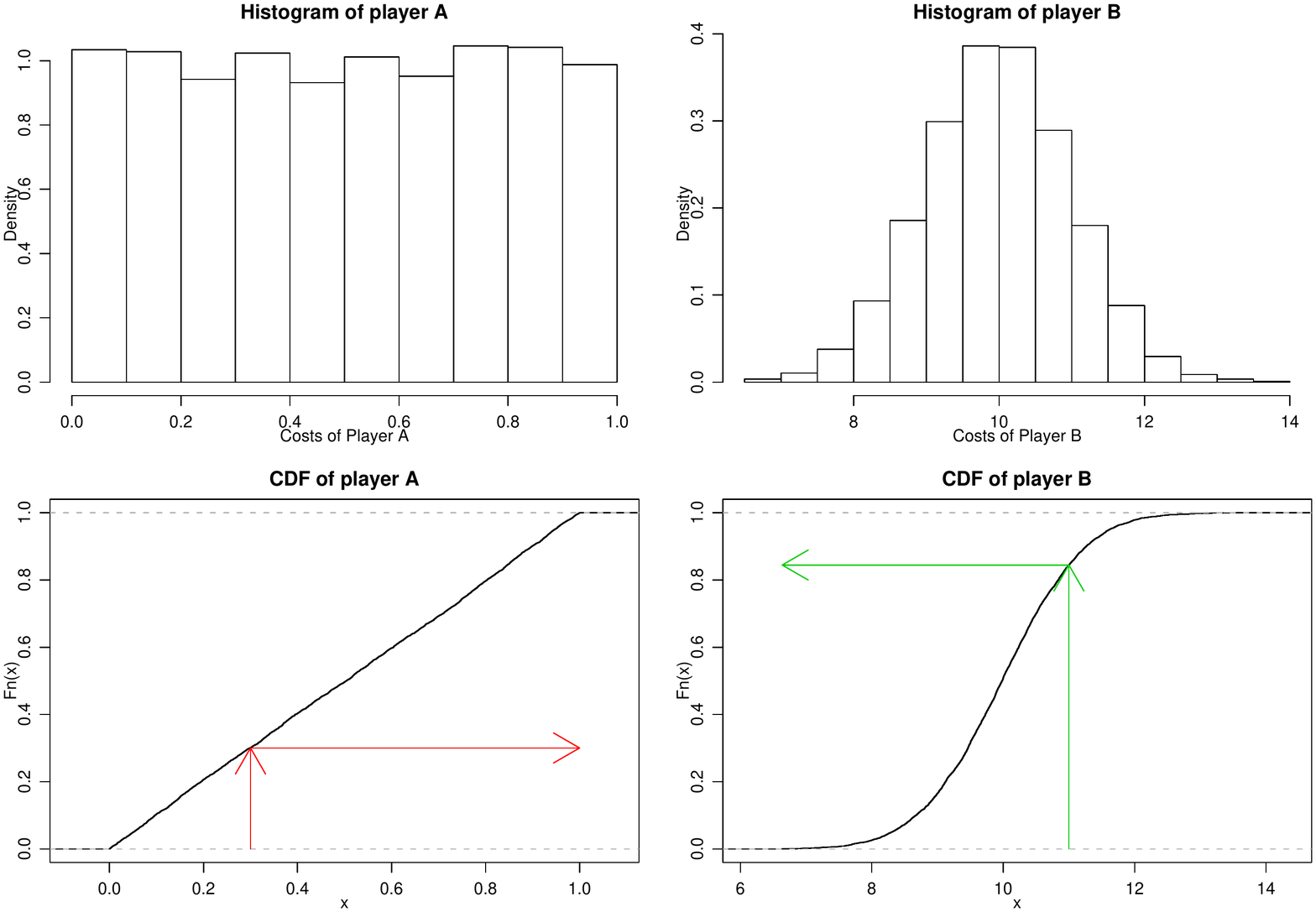}}
    \end{center}
    \caption {At the top, we can see the execution task cost histograms of two different players. Note that they follow different probability distributions. At the bottom, we depict the Cumulative Distribution Function (CDF) for both. As it can be noticed though the depicted arrows, the fact that player $A$ has a minor cost than player $B$  ($0.3$ versus $11$) does not mean that player $A$ will be assigned the task. Instead, when applying the PIT, player $B$ is the one publishing the lower normalized cost.}
     \label{fig:PITUtility}
\end{figure}

Coming back to Algorithm \ref{alg:SimpleLinkingMechanism}, our idea is to modify it by applying the PIT  on the players' declared costs. Hence, instead of publishing the values from its real probability distribution, a player must publish the normalized ones, so that the new algorithm chooses for execution the player minimizing the normalized cost values instead of the original costs. Fig. \ref {fig:PITUtility} illustrates this process. 

Based on these arguments, it is clear that the PIT provides a mechanism for comparing (normalized) node costs. However, we may wonder if the proposed transformation is valid, in the sense that it may not preserve the preferences of the player. To solve this issue, it suffices to notice that, what we are doing is changing the space of preferences. Therefore, the PIT somehow means that, instead of asking the user ``How much does it cost to execute the task?'', we inquire for something like ``What percentage of tasks do you prefer to this one?'' At the end of the day, and for our objectives, these questions are requesting the same information, but the latter is normalized in the interval $[0, 1]$, which is a great advantage. 

Although from an analytic point of view we assume that players could compute the PIT perfectly, in a practical set up players do not need to consider any a priori distribution of probability. They can simply generate costs using their particular distribution and apply the PIT using the successive generated samples.  This process uses what in statistics is known as the Empirical Cumulative Distribution Function (ECDF). We will review this concept later, when we analyze the practical formulation of QPQ in subsection \ref{PracticalQuidProQuo}.

\paragraph{Acceptance test}

Once we know the properties of the PIT, it is clear how we can implement the linking mechanism for the acceptance test. The idea is that any player applying correctly the PIT on her real cost distribution, must generate a uniform distribution on the unit interval on her published normalized cost values. Hence, from the point of view of the mechanism designer, the problem consists on determining whether these published values follow or not that uniform distribution.  There are a wide range of tests that allow checking that. These tests are called \emph{Goodness-of-Fit} (GoF) tests. 

Continuing with this argument, we propose to implement the acceptance test of our algorithm by using some GoF test on the declared transformed sequence of costs published by the player. Whenever a player is honest and she declares the values by applying the PIT transformation on her own distribution, these values will be uniformly distributed in the unit interval. In that case (with high probability) the GoF tests will accept the samples. More important, this process has an error which tends to zero when the number of samples (rounds) increases for any reasonable value of the threshold. For the study of our analytic results, we assume that GoF tests are perfect and this error is zero. (We will review this concept again in our practical implementation of QPQ, in Section~\ref{PracticalQuidProQuo}.)

\paragraph{Punishment}

In the case that a dishonest player tries to avoid the execution of tasks, one possible strategy is to generate increasing cost values, so that the PIT transformed values are close to the unit. However, this type of behavior is quickly detected by the test. An open question is how to establish a punishment to this and any other player whose GoF test comes out negative. One possibility is to force the node to execute the task. Unfortunately, this policy would force fair players to execute tasks in cases of false negatives.  

Another possibility, inspired on previous works on linking mechanisms, is to reject the value declared by the player and generate a new random value according to the normalized uniform distribution. Additionally, we require that no central entity exist on the system. For these reasons, we propose to use a deterministic (repeatable) random generator that any of the remaining nodes can use to calculate the new value. (We deal with the practical aspects of this approach in Section~\ref{PracticalQuidProQuo}.)
At a first sight, this strategy may seem a very poor punishment,  given that there is always a chance that a player emerges victorious of a lie. However, later in this paper we will prove that this is not only enough to discourage dishonest players, but also a crucial ingredient to guarantee that our mechanism is strategy-proof.

\section{The Quid Pro Quo Mechanism}
\label{s:qpq}

After describing the different ingredients of our solution, we are able to propose the final algorithm, which we call the \textit{Quid Pro Quo} (QPQ) mechanism. The details can be observed in Algorithm \ref{alg:RepeatedQuidProQuo}.

\algsetup{indent=2em}
\begin{algorithm}[h!]
\caption{Quid Pro Quo mechanism (code for node $i$, and a generic task $t$, omitted)} \label{alg:RepeatedQuidProQuo}
\begin{algorithmic}[1]
\STATE Estimate the cost  $c_i$ of the task
\STATE Publish the normalized cost $\PITc_i = \mathit{PIT}(c_i)$
\STATE Wait to receive the normalized costs $\PITc_j$ from the other players
\FORALL{$j \in N$ } 
	\IF{not $\mathit{GoF\_Test}(\PITc_j, \mathit{Historic}_j, p\mbox{-}th_j)$}
\STATE $\PITc_j \leftarrow \mathit{Random}(\PITc_{-j})$
\ENDIF
\STATE $\mathit{Historic}_j \leftarrow  \mathit{Historic}_j \cup \{\PITc_j\}$
\ENDFOR
\STATE Let $d = \argmin{j \in N}{\PITc_j}$ .
 \IF{ $d = i$ } 
 \STATE execute the task
  \ELSE 
  \STATE do nothing (node $d$ will execute the task)
  \ENDIF
\end{algorithmic}
\end{algorithm}

Note that we use $\PITc_i$ to denote the PIT-normalized cost to the published, while $c_i$ is the actual cost. We also put in $\PITc_i$ the pseudorandom value that
replaces the value published by $i$ when it does not pass the acceptance test. (Hopefully context will allow disambiguation.) It is important to notice that the algorithm is the same for all participants, and that it is based on information known by them. Therefore, no central entity is required. When a task is issued, each node can estimate its own cost and publish its PIT-normalized value. This value is then received by all other players. When a player has all the values, she checks whether any player published a dishonest value by applying the GoF test. If the value does not pass the test, it is regenerated as described above, by using a pseudorandom generator (that allows all players to generate the same value) of uniformly distributed values in $[0,1]$. With these reviewed values, the player proceeds to determine if its own value is the minimum, in which case it executes the task, publishing the results to the rest of nodes if necessary.

In the following sections, we formally study the expected harm (or reduction of benefit) that dishonest behavior causes on QPQ. Intuition says that the loss due to a dishonest player should be comparable to having that player executing tasks at random.
Indeed, we show below that, independently of their behavior, nodes may never expect a profit of less than the one obtained through a mechanism in which tasks are randomly assigned. This property is very useful in case the node is not capable of accurately evaluating its costs (it is non-rational).

Another important aspect is that QPQ guarantees a minimum benefit to the entire system, even if one or more players are non-rational or rationally limited. In this sense, we will show that the best strategy for any player is to act as if the rest of the players were rational and fair. That is, incorrect behaviors of some players does not alter the strategy of correct players. In next section, we prove all these claims in a formal way.

\subsection{Formal Analysis of QPQ}

Our QPQ algorithm is strongly inspired on the work of Jackson et al.~\cite {LinkingCollectiveDecisionsEfficiency}. Hence, some of our proofs have been adapted from the ones provided there. 
We review now the most relevant properties of the QPQ mechanism presented in Algorithm \ref {alg:RepeatedQuidProQuo}. Assuming that the number of rounds (tasks) is large enough, and that players' costs are independent to each other, we prove the following properties.
\begin{enumerate}
\item QPQ is optimal in the sense that it minimizes the total work done when all players are honest.
\item For any player, the rest of players can be seen as a single aggregated player. For each task, the aggregated player's cost is the smallest of its
members'. These costs follow a Beta distribution.
\item The best strategy of a player is independent of the behavior of the rest.
\item The strategy that optimizes the utility of a player is being honest. In game theory terminology, this means that QPQ is strategy-proof.
\item Each player always obtains a positive expected utility, which is determined by the number of players.
\item An irrational or rationally-limited player always obtains a positive profit.
\item The system is fair in the allocation of tasks and in normalized effort. That is, all the players will run the same number of tasks and perform a similar normalized effort (in expectation).
\item When the number of player is high enough, QPQ ensures very attractive performance.
\end{enumerate}

To address the mathematical analysis of the algorithm we will assume that the PIT and GoF steps are perfect. In fact, with a large number of samples, these processes have errors close to zero.
Another aspect that will simplify our analysis, is the idea of \emph{aggregated player}. We evaluate the performance of a node playing against a ``fictitious'' node that aggregates the responses of all other nodes.  This aggregated player behaves by publishing at each round the minimum of all the normalized costs of the players in the aggregation. This approach is compatible with all the assumptions of the model and is helpful because it significantly simplifies the analysis. 

To make our notation clearer, given a task, we use $x$ to denote the true normalized cost  $\PITc_i$ of player $i$ for that task, while $X$ or $X_i$ is the random variable for that value. When executing QPQ, players may publish $x = \PITc_i$ or another false value. In that case, we use $z$ to denote that dishonest value $\hat{c}_i$ and also, overloading the notation, the re-generated random value replacing it when the GoF test fails. We assume that the $z$ values are realization of some random variable $Z$. Given a task with cost $c_i$, the player obtains a normalized utility $\PITu_i=\PITc_i$ when she does not execute the task (independently on what she published) and makes a normalized work of $\PITw_i=\PITc_i$ when she executes the task (where $\PITW_i$ denotes the random variable). Additionally, we use $y$ to denote the value $\min{\PITc_{-i}}$ published by an aggregated player. Following mechanism design notation, we say that the (social) decision function $d$ of QPQ is  
$$ d = \argmin{i \in N} {\PITc_i}.$$ 
Then, we define $\mathit{Pr}[ d = i]$ as the probability that player $i$ declares the minimum value and executes the task. When working with the aggregated player,  $Y$ is a vector of random variables, and we use $\mathit{Pr}[Y \leq y]$ to denote the probability that at least one element of $Y$, say $j$,  validates $Y_j \leq y$. 

With this notation in mind, we can prove that, for any player $i$, the expectation of the declared costs is equal to the expected utility plus the expected work. Additionally, this quantity is a constant. I.e.,
\begin{equation}
\label{umasw}
\mathit{E} [\PITC_i] = \mathit{E} [\PITU_i + \PITW_i] = \mathit{E} [\PITU_i] + \mathit{E} [\PITW_i] = \int_0^1 \PITc_i d \PITc_i = \frac{1}{2}.
\end{equation}
This means that a player maximizes her utility when she minimizes her work, and vice-versa. In the following propositions, we will use this fact.

\paragraph{Players' normalized costs distributions}

We argue here that all players' normalized costs follow independent uniform distribution on $ [0,1] $.
When players are honest, their report values follow a uniform distribution on $ [0,1] $. This follows from the properties of the PIT transformation introduced above. On the other hand, when a player is dishonest, it may change the distribution of its normalized costs trying to obtain extra benefit. However, we assume that in this case the GoF test fails. Then, her attempt will be detected, and the value will be replaced by pseudorandom value
drawn from an independent uniform distribution on $ [0,1] $. A final case is that the dishonest player may generate fake normalized cost that follow a uniform distribution on $ [0,1] $, hence passing the GoF test. In this case the normalized cost $\PITc_i(t)$ for a task $t$ is independent from the values $\PITc_j(t)$ of the other players, since from concurrency
the value has to be sent before the others are received.
Hence, the following result.

\begin{prop}
\label{p:iid}
The set of final normalized costs considered in Line 10 of Algorithm~\ref {alg:RepeatedQuidProQuo} are drawn from independent and identical distributed (iid) random variables, with uniform distribution on $ [0,1] $.
\end{prop}

\paragraph{Optimality}
The QPQ algorithm is optimal in the sense that, if all players are honest, it minimizes the total normalized work done. 
\begin{prop}\label{eq:optimality}
Assume that all players are honest. For a given set $T$ of tasks, there is no mechanism $M$ such that
$\mathit{E}[\sum_{i = 1}^n \PITW^M_i] < \mathit{E}[\sum_{i = 1}^n \PITW_i]$, where $\PITW^M_i$is the random variable associated with the normalized work done by player $i$ when using mechanism $M$.
\end{prop}
\begin{proof}
The proof is straightforward using contradiction. Assuming that such mechanism $M$ exist, there must be, at least, one task for which $\PITw^M < \PITw$, however, the social decision function of QPQ always selects the player publishing the minimum of the normalized costs, so it is not possible that $M$ is able to select another player capable of executing with less cost. 
So, we conclude that $M$ cannot exist.
\end{proof}

\paragraph{Aggregated player}

It is assumed that players' normalized costs have independent uniform distributions on $ [0,1] $. Hence, the probability density function of each player $ i $ is $ f_i (\PITc_i) = 1 $ on that interval.
Thus, the costs of an aggregate player for $n-1$ nodes follows a probability distribution $\mathit{Beta} (y; 1, n-1) $ as shown.

\begin{prop}
The costs $Y$ of the aggregated player of $n -1$ i.i.d.~players (with uniform distribution on $ [0,1] $) follows a $Beta(y; 1, n-1)$ distribution, with density probability function $f(y) = (n-1) (1-y)^{(n-2)}$ and CDF $F[y] = \mathit{Pr}[Y \leq y]=1 - (1 - y)^{n-1}$. \label{AgregatePlayer}
\end{prop}
\begin{proof}
Recall that the cost of an aggregated player is the minimum of the normalized costs of the players in the aggregation. The CDF $F(\cdot)$ of that cost can be obtained as follows. Let us assume that the players in the aggregation are 1 to $n-1$.
\begin{align}
F[y] & = \mathit{Pr}[Y \leq y] = 1 - \mathit{Pr}[Y > y] \\
 & = 1 - \prod_{j = 1}^{n -1} \mathit{Pr}[Y_j > y] \\ 
 & = 1 - \left(\int_y^1 1 d \varphi \right)^{n-1} \\
 & = 1 - (1 - y)^{n-1}
\end{align}
Where $Y_j$ is the random variable associated with the normalized cost of node $j$. Hence, the density probability distribution is
$$
f(y) = (n-1) (1-y)^{(n-2)}.
$$
The $\mathit{Beta}$ distribution is  defined as follows \cite{StatisticalDistributions2000}.
\begin{align}
Beta(y; 1, n-1) & = \frac{1}{B(1,n-1)} y^{1-1} (1 - y)^{(n - 1) - 1}\\
 & = \frac{(1 + (n - 1) - 1)! }{((n - 1) - 1)!} (1 - y)^{n - 2},
\end{align}
where $B(\cdot,\cdot)$ is the Beta function. Now, it is easy to check that $f(y) = \mathit{Beta}(y; 1, n-1)$.
\end{proof}

\paragraph{Players' strategies}

Every rational player knows that the rest of players follow uniform and independent distributions. The question a selfish rational player makes is which is the best strategy for obtaining the greatest possible benefit. If a player uses a distribution other than the uniform,
her values will be rejected by the GoF, and will be re-generated from a uniform distribution. 
However, a player could lie following a uniform distribution that is not independent of her actual values. Note that QPQ does not know about true normalized costs (they are private) and uses for the assignment decision the declared value or the random value assigned by the system if a lie is detected. In both cases, the aggregated player see a random variable $Z$ that must follow a uniform distribution.
We show now that either case drives the player to worse results that her own honest distribution, then that player will no have any incentive to cheat.

Let us first quantify the expected work done by honest players.

\begin{prop}\label{eq:utilitysavedwork}
The expected normalized work $E [\PITW_i]$ done by an honest player $i$ is $\frac{1}{n+n^2}$.
\end{prop}
\begin{proof}
Recall that we assume that player $i$ is in the system with an aggregated player of $n-1$ nodes. Then,
$\mathit{Pr}[ d = i]$ is the probability that player $i$ publishes a normalize cost smaller than the one of the aggregated player.
\begin{align}
\mathit{E} [\PITW_i] & = \int_0^1 x \mathit{Pr}[ d = i] dx \\
& = \int_0^1 x \int_x^1 (n-1) (1-y)^{(n-2)} dy dx \\ 
& = \frac{1}{n+n^2}.
\end{align}
Notice that we use the probability distribution of the aggregated player derived in Proposition~\ref{AgregatePlayer}.
\end{proof}



\begin{prop}\label{eq:samecontraryutility} The total normalized work done by an aggregate player $j$ (aggregating $n-1$ nodes), with costs $x=\PITc_j$, does not change when a player $i$ (not in the aggregation) declares dishonest values $ z = \hat{c}_i$. 
\end{prop}
\begin{proof}
Let us abuse the notation and use $z$ to denote the dishonest values declared by $i$ if the GoF is passed or the re-generated
values if it does not. Let $Z$ be the uniform random variable associated with these values.
We assume that there is a bi-variate probability distribution with density $ f_ {x, z} (x, z) $ that relates both values $x$ and $z$.
In that case, the marginal distribution for $z$ must be uniform.  Therefore, we have,
\begin{align}
f_x(x) = \int_0^1 f_{x,z}(x, z) dz = 1 \\
f_z(z) = \int_0^1 f_{x,z}(x, z) dx = 1 \\
\int_0^1 \int_0^1 f_{x,z}(x, z) dz dx = 1.
\end{align}
Hence, the expected work done by $j$ is
\begin{align}
\mathit{E} [\hat{W}_j] &= \int_0^1 y  \mathit{Pr}[ d = j ] dy \\
& = \int_0^1 y (n-1) (1-y)^{(n-2)} \int_0^1 \int_y^1 f_{x,z}(x, z) dz dx dy \\
& = \int_0^1 y (n-1) (1-y)^{(n-2)}  \int_y^1 \int_0^1 f_{x,z}(x, z) dx dz dy,
\end{align}
where $\mathit{E} [\hat{W}_j]$ is the expected work done by the aggregated player $j$ when player $i$ lies. But, as we have uniform marginals, the above expression becomes
\begin{align}
\mathit{E} [\hat{W}_j] &= \int_0^1 y (n-1) (1-y)^{(n-2)} \int_y^1 1 dz dy \\
& = 
 \int_0^1 y (n-1) (1-y)^{(n-2)}(1-y) dy \\
 & = \int_0^1 y (n-1) (1-y)^{(n-1)} dy \\
 & = \frac{n-1}{n+n^2}.
\end{align}
Which is equal to the total wok done by the aggregated player $j$ when $i$ is honest, that can be computed as follows.
\begin{align}
\int_0^1 y (n-1) (1-y)^{(n-2)} \int_y^1 1 dx dy 
& = 
\int_0^1 y (n-1) (1-y)^{(n-1)} dy \\
& = \frac{n-1}{n+n^2}.
\end{align}
\end{proof}

In summary, an aggregate player $j$ expects to performs the same amount of work, independently of the behavior of a given player $i$ not in the aggregation.
I.e., its expected work is not affected by whether $i$ is honest or dishonest. This allows us to prove that the optimal strategy for a player is to be honest.

\begin{prop} 
A player $i$ never does more normalized work (in expectation) by being honest. That is,
\begin{align}\label{eq:bestutility}
 \mathit{E}[\PITW_i] \leq  \mathit{E} [\hat{W}_i],
\end{align}
where $\mathit{E} [\hat{W}_i]$ is the expected work performed by player $i$ when it is dishonest.
\end{prop}
\begin{proof} 
For the sake of contradiction, let us suppose this proposition is false. Hence, there is some set of tasks for which, if $i$ is
not honest, it performs less work in expectation. I.e., $ \mathit{E} [\PITW_i] >  \mathit{E} [\hat{W}_i]$.
Additionally, using Proposition~\ref{eq:samecontraryutility}, we know that the aggregated player, $j$, will do the same expected work, i.e., $ \mathit{E} [\PITW_j] =  \mathit{E} [\hat{W}_j]$.
Hence, it follows that 
$ \mathit{E} [\PITW_i] +  \mathit{E} [\PITW_j] > \mathit{E} [\hat{W}_i] +  \mathit{E} [\PITW_j]$.
However, if the above inequality were true, QPQ would not be optimal, since a mechanism that reproduces the same task assignments done under $i$ lying (in presence of honest players would have less expected work). Clearly, this is in contradiction of Proposition~\ref{eq:optimality}.
Therefore, the best strategy for a player (the one minimizing her normalized work done) is to be honest.
%
%
\end{proof}

We complement this result with the following property.
\begin{prop} 
When a player $i$ publishes dishonest non uniform values or values independent of her true normalized uniform distribution, it performs in expectation $\mathit{E} [\hat{W}_i] = \frac{1}{2 n}$ work.
\end{prop}
\begin{proof}
The values $z$ used to decide whether to assign a task to player $i$ follow a uniform distribution that is independent of the actual costs for $i$. Hence,
\begin{align}\label{eq:utilitydiffvalue}
\mathit{E} [\hat{W}_i] & = \int_0^1 x \mathit{Pr}[ d = i] dx \\
& = \int_0^1 x  \int _0^1 \int_z^1 (n-1) (1-y)^{(n-2)} dy dz dx \\  
& = \frac{1}{2 n}
\end{align}
%
\end{proof}

From this result, Proposition~\ref{eq:utilitysavedwork}, and Eq.~\ref{umasw}, we directly derive the following theorem.

\begin{thm}
Given that $n \geq 2$, it holds that $E [\PITW_i]=\frac{1}{n+n^2}  <  \frac{1}{2 n}=\mathit{E} [\hat{W}_i]$. Hence, since
the sum of the expected work and expected utility is $\frac{1}{2}$, players obtain higher expected utilities by being honest than by publishing dishonest normalized costs.
\end{thm}

\paragraph{Real expected utility}
Note that the normalized work done by a honest player, as calculated above, is equal to $ \frac {1} {n +n^2} $. But we may wonder what is the real (not normalized) work done. We can easily calculate it in terms of real utility as follows.

\begin{thm}[] For each player $i$, the real expected utility is 
$$\mathit{E}[U_i] = \int_\Omega {x f_i(x) (1 - (1 - F_i(x))^{n-1}) dx}. \label{trueUtilityWithN}$$
where the real cost of player $i$ is a continuous random variable with support $\Omega$, probability density function $f_i(\cdot)$, and CDF $F_i(\cdot)$.
\end{thm}
\begin{proof}
Let $Y_j = PIT(X_j) = F_j(X_j)$ be the uniform random variable that gives the normalized cost for player $j \neq i$ at the time of assigning the tasks (Line 10), and
$Y=\min_{j\neq i}{\{Y_j\}}$. 
Then, the expected (real) utility of player $i$ is: 
\begin{align*}
\mathit{E}[U_i] & = \int_\Omega {x f_i(x)\mathit{Pr}(Y \leq F_i(x)) dx} \\
& = \int_\Omega {x f_i(x) (1 - (1 - F_i(x))^{n-1}) dx},
\end{align*}
where we have used that $\mathit{Pr}[Y \leq y]=1 - (1 - y)^{n-1}$ (Proposition~\ref{AgregatePlayer}).
\end{proof}

\subsubsection{Fairness}

The following result, combined with Proposition~\ref{p:iid}, will be used to show that all players execute, on expectation, the same number of tasks, even when some players are non-rational or dishonest. 

\begin{prop}
Let $X_1, X_2, \ldots, X_n$ be $n$ continuous and i.i.d.~random variables, then:
$$\mathit{Pr}(X_i \leq \min_{j \neq i} {\{X_j\}}) = \frac{1}{n}. \label{probXlessYwithN}$$
\end{prop}
\begin{proof}
\begin{align*}
\mathit{Pr}(X_i \leq \min_{j \neq i} {\{X_j\}})
& = \int_{-\infty}^\infty f(y) [ \int_{-\infty}^y f(x) dx ]^{(n-1)} dy \\
& = \int_{-\infty}^\infty f(y)  [F(y)]^{(n-1)} dy \\
& = \int_{-\infty}^\infty [F(y)]^{(n-1)} dF(y) \\
& = \left[\frac{\{F(y)\}^{n}}{n} \right]_{-\infty}^\infty \\
& = \frac{1}{n}
\end{align*}
\end{proof}

Hence, QPQ not only offers best utility guarantee to honest rational players, but it also offers good properties in environments where nodes have difficulty in estimating costs. This is because, even in environments where the nodes are non-rational, QPQ divides the work fairly and optimaly with respect to the declared normalized costs.
Clearly, non-rational players run major efforts, but it is always under completely random task assignments. In other words, the extra cost of non-rational players is caused by their own ignorance, not by the wickedness of the other players.
Then, given that players are are assigned tasks by choosing the smallest value from a set of i.i.d.~random variables (Proposition~\ref{p:iid}), QPQ ensures that the expectation of the number of tasks executed by each node is $ |T| / n $, where recall that $T$ the set of tasks and $n$ the number of players. 

\begin{cor}
In QPQ, players will execute in expectation  a proportion of $\frac{1}{n}$ of the tasks, and thus a total of $\frac{|T|}{n}$ of tasks.
\end{cor}
\begin{proof}
Declared values follow continuous and identically distributed (uniform) random variables in $[0,1]$, from Proposition~\ref{p:iid}, and therefore applying  Proposition \ref{probXlessYwithN}, each player will execute in expectation a proportion of $\frac{1}{n}$ of the tasks.
\end{proof}


 

\subsubsection{Bounds}

Finally, we think that it could be interesting to define some ratio that measures how the efficiency of the QPQ mechanism degrades with the selfish behavior of the players. Following concepts similar to the ``price of anarchy'' \cite{KoutsoupiasPapadimitriouWorstcaseEquilibria}, we define the measure of \emph{efficiency} as the ratio between the utility of an equilibrium (usualy the ``worst equilibrium'') and the utility of some optimal solution.

Obviously, the player's normalized utility must be between $ 0 $, when the node runs all tasks, and $ \frac {1} {2} $ when the node has not executed any task. But there are two levels that may be considered as references to establish the goodness of the algorithm. On one side, when a node runs completely random $\frac {1} {n}$  tasks, the expected effort is $ \frac {1} {2n} $. On the other hand, the maximum benefit a player $i$ could get occurs when its tasks correspond exactly to her cheapest tasks. In this case, the expected utility would be
\begin{align}
\mathit{E} [\PITU_i^*] & =  \frac {1} {2} - \mathit{E} [\PITW_i^*] \\
& = \frac {1} {2} - \int_0^\frac {1} {n} x dx \\
&=  \frac {1} {2} - \frac {1} {2n ^ 2}.
\end{align}

Although this case has null probability, we propose to use this concept for our definiton of measure of efficiency.

\begin{defn}[Measure of efficiency].  
We define the measure of \emph{efficiency} of an algorithm $M$ for tasks assignment under selfish behaviour as the ratio between the expected normalized utility obtained under some equilibrium and $\mathit{E} [\PITU_i^*]= \frac {1} {2} - \frac {1} {2n ^ 2}=\frac {n^2-1} {2n ^ 2}$. I.e.,
$$\text{Efficiency} =\frac{\mathit{E}[\PITU^M]}{\mathit{E} [\PITU_i^*]} = \frac{{2n ^ 2} \mathit{E}[\PITU^M]}{{n^2-1} }.$$
\end{defn}


Hence, we can compute the efficiency of QPQ as
$$ \text{Efficiency} = \frac{2n ^ 2 (\frac {1} {2} - \frac{1}{n+n^2})}{n^2-1}
= \frac{n ^ 2 - (\frac{2n}{n+1})}{n^2-1} > \frac{n ^ 2 - 2}{n^2-1}
.$$
Note that the efficiency of QPQ is close to $1$ when the number of participants is high. For instance, with just $10$ nodes the efficiency of QPQ is $0.991$.

\section{Implementing QPQ in real environments} 
\label{PracticalQuidProQuo}

In this section, our objective is analyzing what are the restrictions for QPQ to be implemented in real environments. From above sections, we may claim that the computation and communication capabilities required by the algorithm are affordable with current technology.
We do not claim that implementing such capabilities would be an easy task, since there are many technological challenges that should be addressed to do it. Other previous works show some of them \cite{Murray:2010:CCC:1851322.1851334}. Thus, our only claim is that it would be feasible.

However, going beyond the required communication and computation capabilities, we may see that a number of issues arise. The first of all is on the definition of selfishness itself. This paper is mainly focused on detecting and neutralizing users publishing values not coming from the PIT of their real costs. However, one can claim that other non-cooperative harmful behaviors are possible such as, for example, not executing tasks at all, or executing them incorrectly. Hopefully, most of these evil conducts can be easily avoided using a two step scheme. First, by detecting such behaviors (previous works on the area show that it is possible \cite{DBLP:conf/ipps/AntaGM10,DBLPjournalspplAntaGLS12}). Second, by establishing a strong enough punishment to discourage misbehaving players from repeating them. For example, we may adopt the radical solution of just sending off misbehaving users. In order to guarantee that reoffending players participate again, all that is needed is that users identities are unique and cannot change on different game instances. 
Note that QPQ does not discard misbehaving users, because it assumes that the publication of dishonest values cannot be distinguished from the publication of values generated from rationally-limited players, and it would not be reasonable to send off the latter from the game given that, in a realistic scenario, all players would have some rationally limitations (i.e. it is not possible to estimate costs with total accuracy). Hence, QPQ's approach of keeping them in the system is one of the most difficult ways of dealing with selfish users.

Coming back to the subtleties of QPQ, another point to consider is how to re-generate the random value when the system detects a lie. As we said before, we require a deterministic (repeatable) random generator that any of the remaining nodes can use to calculate the new value. One possibility for generating the random value is to use a hash function over the published normalized cost of other nodes. Alternatively, it is possible to request a random value to each player (except the value of player in question) and apply the hash function on them. Even another possibility is to use techniques similar to the procedures proposed by Aumann et al. \cite {aumann1995repeated} to generate jointly controlled lotteries. For example, for two players, we can request random values to both, and replace the value of the liar's by the sum of these numbers, if the sum is less than $1$, or with one minus the sum otherwise. With this scheme, it is easy to show that when one of the player declares random values according to a uniform distribution, then this process generates random values also uniformly over $[0, 1]$, regardless of what the other player does. As a conclusion, we may claim that there are several mechanism suitable for the generation of the punishment random value independently on the behavior of a dishonest player.

Another obstacle that stands on the way of a potential implementation of the mechanism is the acceptance test. We have assumed that we have a perfect GoF test function. This is somewhat similar to assume that we have a set of samples whose number is very large (ideally infinite) for detecting lies with the usual tests. In a real system, this solution is impractical since nodes would require to store all the historical values of the rest of players, and initially the number of samples is necessarily limited. As we saw before, we propose that players simply generate costs using their particular distribution and apply the PIT using the successive generated samples. The CDF used for the PIT is synthesized from the existing samples $y_i$. This CDF obtained from samples is known in statistics as the Empirical Cumulative Distribution Function (ECDF). 

\begin{defn}[Empirical Cumulative Distribution Function]
\label{ECDF} The empirical cumulative distribution function (ECDF)
$F_n$ for $n$ observations $y_i$ is defined as
    $$F_n(x)={\frac {1}{n}}\sum_{i=1}^n  \mathbf{1} \{ y_i\leq x \},$$
    where  $\mathbf{1}\{A\}$ is the indicator function or the characteristic function of event $A$. In our case, it is defined as
    $$\mathbf{1}\{ y_i\leq x \} = \left\{\begin{matrix}1 & \mathrm{if}\ y_i\leq x, \\ 0 &
    \mathrm{otherwise}.\end{matrix}\right.$$
\end{defn}

Obviously, this process has an error which tends to zero when the number of samples (rounds) increases as it is proved by \emph {Glivenko-Cantelli theorem} \cite{GlivenkoCantelliTheorem}.

Regarding the GoF used, a tremendous number of GoF tests have been proposed in the scientific literature. Some of them may be applied over discrete distributions and others require continuous distributions. The Kolmogorov-Smirnov (KS) test \cite{KolmogorovAN_1933,Smirnov_NV_1939} is probably the best-known test for continuous distributions, basically due to its simplicity. The KS test calculates the greatest distance between the ECDF associated to a sequence of samples and the CDF we want to check. It may be defined by the following expression:
$$D = \max_{1\leq i\leq n}(F(x_i)-\frac{i-1}{n}, \frac{i}{n} -F(x_i))$$
where $F(\cdot)$ is the CDF to check, $n$ is the number samples, and $(x_1, x_2, \cdots, x_n)$ is the set of samples arranged in increasing order.
What makes the KS test so versatile is that the distribution of the distance $D$ does not depend on the theoretical probability distribution (null hypothesis). Several authors, such as Smirnov \cite {Smirnov_NV_1939}, Birnbaum and Tingey \cite {birnbaumtingey1951}, have obtained exact and approximate expressions of the distribution of the variable $D$ as a random function of the number of available samples. Due the complexity of such expressions, the KS test is often used through tables containing the most common percentiles. 

\begin{figure*}[ht]
  \centerline{\includegraphics[width=16cm]{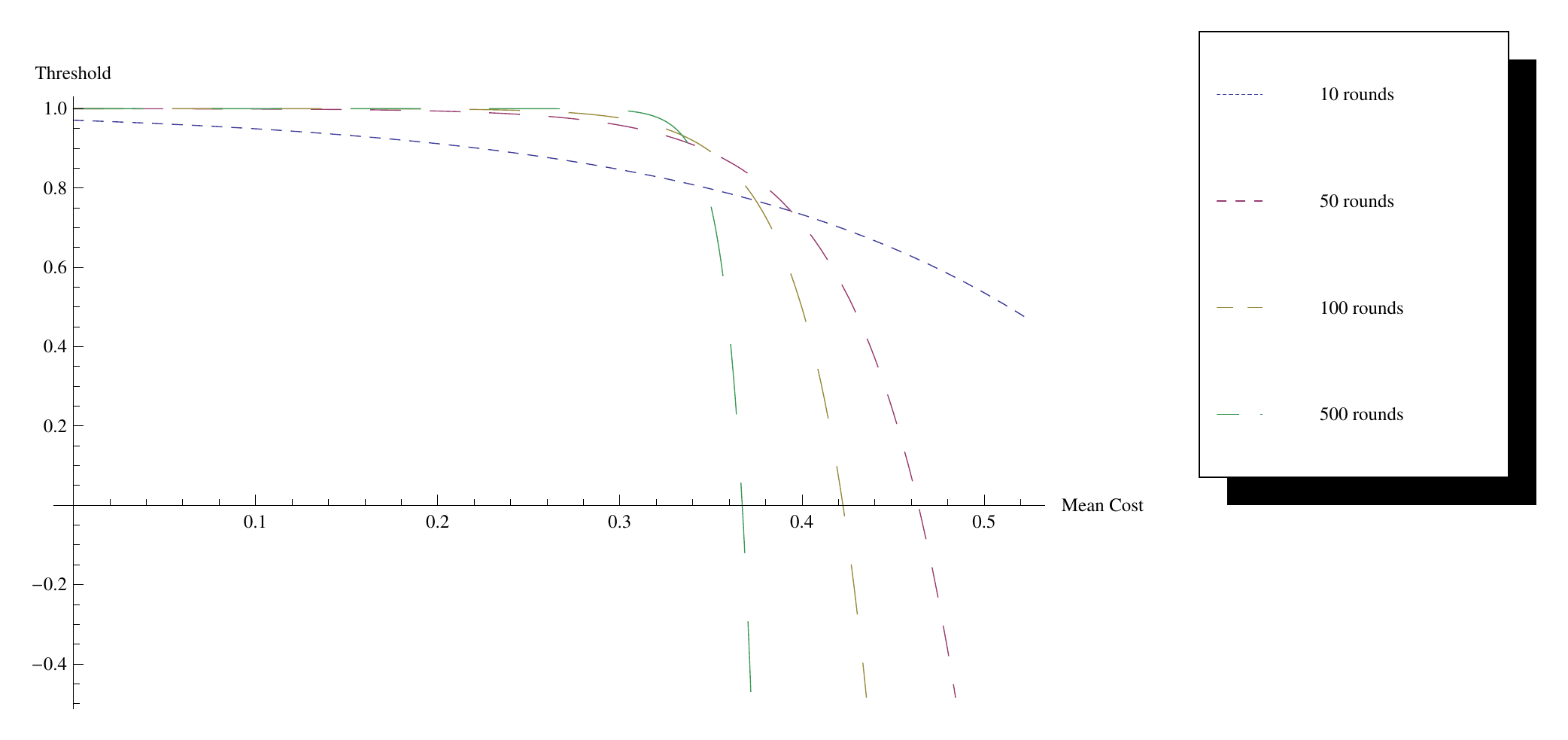}}
  \caption{
  This picture depicts the curves of our elastic \emph{p-value} thresholds as function of the normalized utility of a player. When we have a small number of rounds (blue line for 10 rounds) our system is quite tough, but if the number of rounds increases (yellow line correspond to 100 rounds and green line is for 500 rounds), our proposal is more relaxed, and accepts values if the player's utility is within a reasonable range.}
  \label{fig:PVALUE}
\end{figure*}

We propose to use the KS test as the GoF test of QPQ. Hence, whenever a new normalized cost is issued, we check the KS test of it, together with the historical sequence of that player, so that we obtain the corresponding (\emph {p-value}). Note that, in statistical significance testing, the \emph{p-value} is the probability of obtaining a particular test statistic on the model at least as extreme as the one that was actually observed. Now, the value is accepted by the test when that \emph {p-value} is over a particular acceptance threshold, $p\mbox{--}th$.


For practical reasons, we need to reduce the history of a user to a relatively small number of samples. 
Hence, we propose a slight modification to the acceptance test of Algorithm~\ref{alg:RepeatedQuidProQuo} to make it implementable in real systems. With this modification, each node applies the KS test using only a small number of the latest published values. 
However, this makes the KS test susceptible of generating inaccurate estimations. For example, a selfish node could publish values following a Beta distribution $(1, 0.9)$. With high probability, this situation could not be detected with sample sequences of small length.
In addition of choosing a large enough sample size (our simulations show that $50$ samples are enough), we play with the threshold to refine the test.
The idea is to modify the acceptance threshold so that it is hardened when the actual normalized utility of the player is higher than the theoretical expectation, and it is relaxed when players are losing more than expected. There are many ways of implementing this idea, but we propose the following expression
 $$p\mbox{-}th_k = \frac {1}{\log(k+1)^{\delta * (1 - (\mu_k - \mu) \cdot \sqrt{k} )}},$$
where $\delta$ is a tuning parameter, $\mu$ is the expected normalized utility of all players and $\mu_k$ is the actual normalized utility of the player at round $k$. To illustrate this idea we depict Fig. \ref{fig:PVALUE}, which represents the value of this threshold as a function of the total normalized utility of the player. Clearly, the above formula is entirely empirical, although the simulations below in this paper show that it fits well our requirements. 
One of the reasons that has led to the development of this proposal has been the idea that a new player must ``pay'' some kind of ``fee'' when she enters into the system. In this way, we want to avoid, or at least reduce, the problem of low-cost identities or cheap pseudonyms. With our proposal, at the beginning QPQ asigns tasks almost randomly, while later, when we have more information about players, QPQ assigns tasks optimally. Each player has to ``pay'' at the beginning working in random assigments and thus, she has no incentive to exit and reenter into the system. 


The final implementable QPQ algorithm we propose may be written as presented in Algorithm~\ref{alg:ImplementableQuidProQuo}.

\algsetup{indent=2em}
\begin{algorithm}[h!]
\caption{Implementable Quid Pro Quo mechanism (code for node $i$, and a generic task $t$, omitted)} \label{alg:ImplementableQuidProQuo}
\begin{algorithmic}[1]
\STATE Estimate the cost  $c_i$ of the task
\STATE Publish the normalized cost $\PITc_i = \mathit{PIT}(c_i)$
\STATE Wait to receive the normalized costs $\PITc_j$ from the other players
\FORALL{$j \in N$ } 
	\STATE Let $p\mbox{-}th_j \Leftarrow \frac {1}{\log(k+1)^{\delta (1 - (\mu_{j,k} - \mu)\cdot \sqrt{k} )}}$
	\IF{not $\mathit{KSTest}(\PITc_j, \mathit{Historic}_j, p\mbox{-}th_j)$}
\STATE $\PITc_j \leftarrow \mathit{Random}(\PITc_{-j})$
\ENDIF
\STATE $\mathit{Historic}_j \leftarrow  \mathit{Historic}_j \cup \{\PITc_j\}$
\ENDFOR
\STATE Let $d = \argmin{j \in N}{\PITc_j}$
 \IF{ $d = i$ } 
 \STATE execute the task
  \ELSE 
  \STATE do nothing (node $d$ will execute the task)
  \ENDIF
\FORALL{$j \in N$ } \STATE recompute $\mu_{j,k+1}$\ENDFOR
\end{algorithmic}
\end{algorithm}

\subsection{Simulations}

\begin{figure}[h]
\centerline{\includegraphics[angle=270,width=7cm]{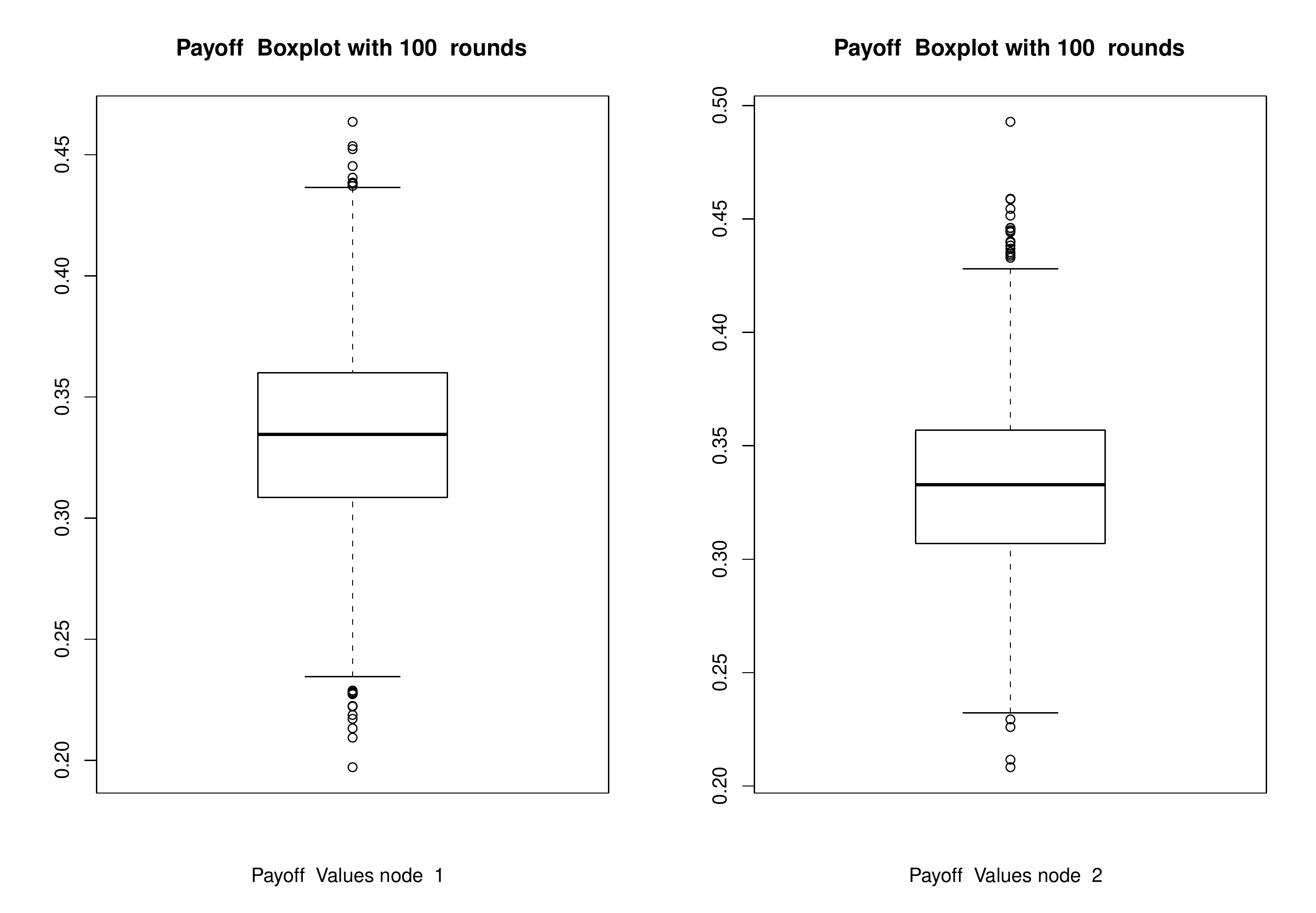}}
\caption{Two honest players with no control (100 rounds).}
\label{fig:SimulationResult2Players100NoControlUnifUnif}
\end{figure}
\begin{figure}[h]
\centerline{\includegraphics[angle=270,width=7cm]{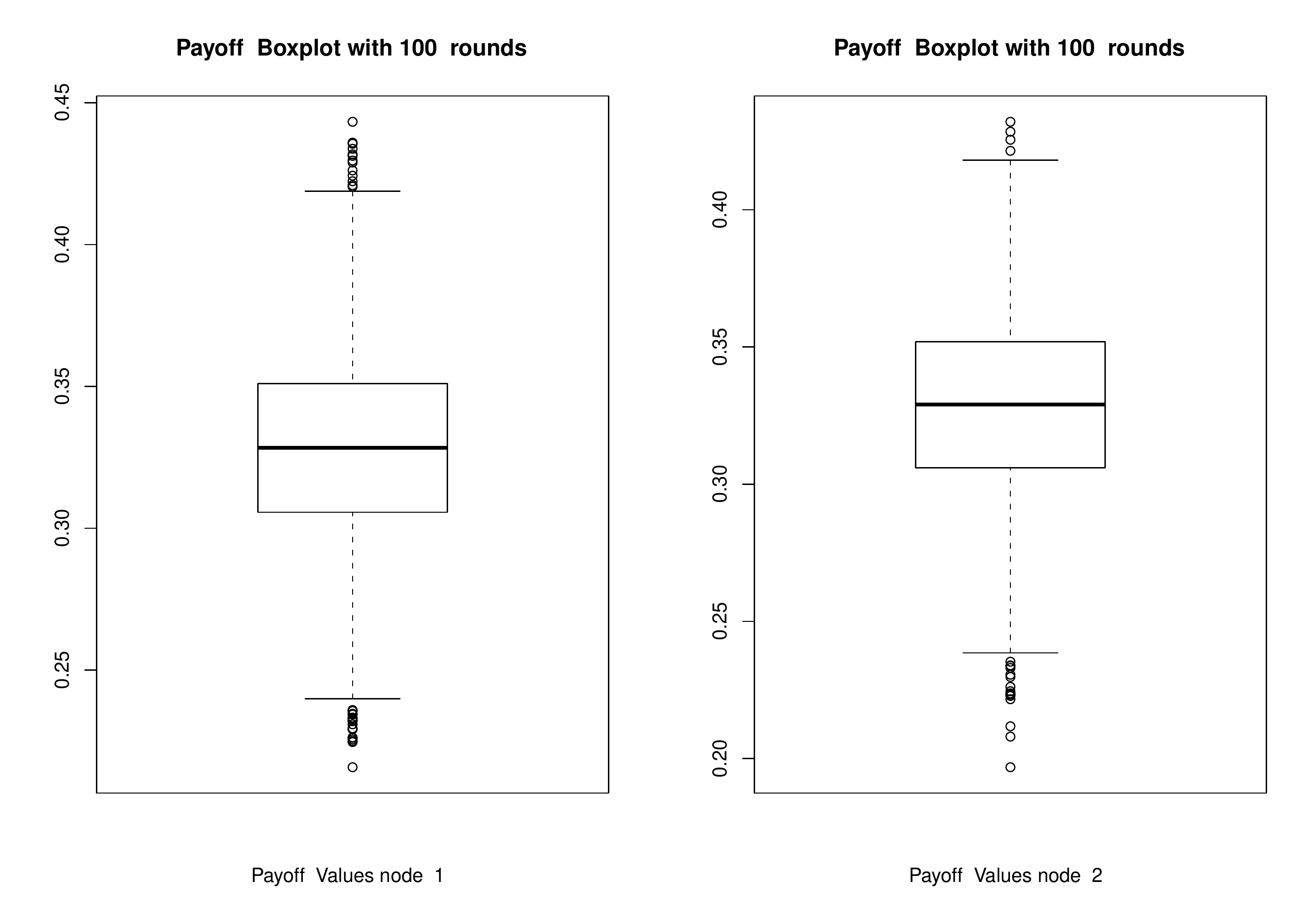}}
\caption{Two honest players with KS control (100 rounds).}
\label{fig:SimulationResult2Players100ControlUnifUnif}
\end{figure}

By performing simulations, we have checked various aspects of the implementable QPQ. First of all, we wondered if the new GoF test may punish fair players by generating false negatives. In this direction, Fig.~\ref{fig:SimulationResult2Players100NoControlUnifUnif} represents the boxplot of the expectation of the normalized work done in 100 rounds when all players are honest and no GoF test is applied. This picture serves as control and allows us to compare it with the same game but introducing the GoF test of Algorithm \ref{alg:ImplementableQuidProQuo}, using a history of $50$ samples for the KS test and $\delta=2$. The results are depicted in 
Fig.~\ref{fig:SimulationResult2Players100ControlUnifUnif}. As it can be seen, the performance loss caused by false negatives is minimal and barely noticeable in these scenarios.


The next question is to which extend selfish users can fool the algorithm and achieve improvements in their utility. We have simulated dishonest behavior by using several distributions close to the uniform but with higher mean, by taking advantage of the properties of the Beta function, so that these distributions try to pass the implementable KS test and, at the same time, obtain some profit on the long run.
%
Again, we have run simulations considering a game with two nodes, one honest (uniform) and one dishonest for a set of $1,000$ rounds, with historical lengh of $50$ samples for the implementable KS test and with  $\delta = 2 $.
The results can be seen in Table \ref {table-payoffs}, which depicts the normalized player utilities for different scenarios. In the table, the name \emph{Uniform} represents honest nodes, \emph{Random} is used for non-rational players generating random costs and finally, \emph {"Beta"} and \emph {"Normal"} are used for dishonest players following those distributions. As it can be observed, honest utilities remain quite constant, while non-rational and dishonest utilities decrease, although never under a given limit. Interestingly note that this behavior is maintained even in the extreme case of a $ Beta (1, 0.9) $ distribution. Observe that, when the number of samples is small (around $50$), a $ Beta(1, 0.9)$ is so similar to a uniform distribution that it is hardly distinguishable to the eye. 

\begin{table}
\begin{center}
\framebox[3.0in]{\begin{tabular}{c || c | c } Distributions & $\PITU_{1}$
& $\PITU_{2}$
\\ \hline
Uniform vs. Uniform & 0.332 & 0.332 \\
Uniform vs. Random & 0.331 & 0.250 \\
Uniform vs. Beta(1, 0.9) & 0.321 & 0.258 \\
Uniform vs. Beta(1, 0.7) & 0.315 & 0.264 \\
Uniform vs. Normal & 0.352 & 0.250 \\
\end{tabular}}
\caption{Honest vs. dishonest utility/cost.} 
\label{table-payoffs}
\end{center}
\end{table}

\begin{figure*}[ht]
  \centerline{\includegraphics[width=9cm, angle=270]{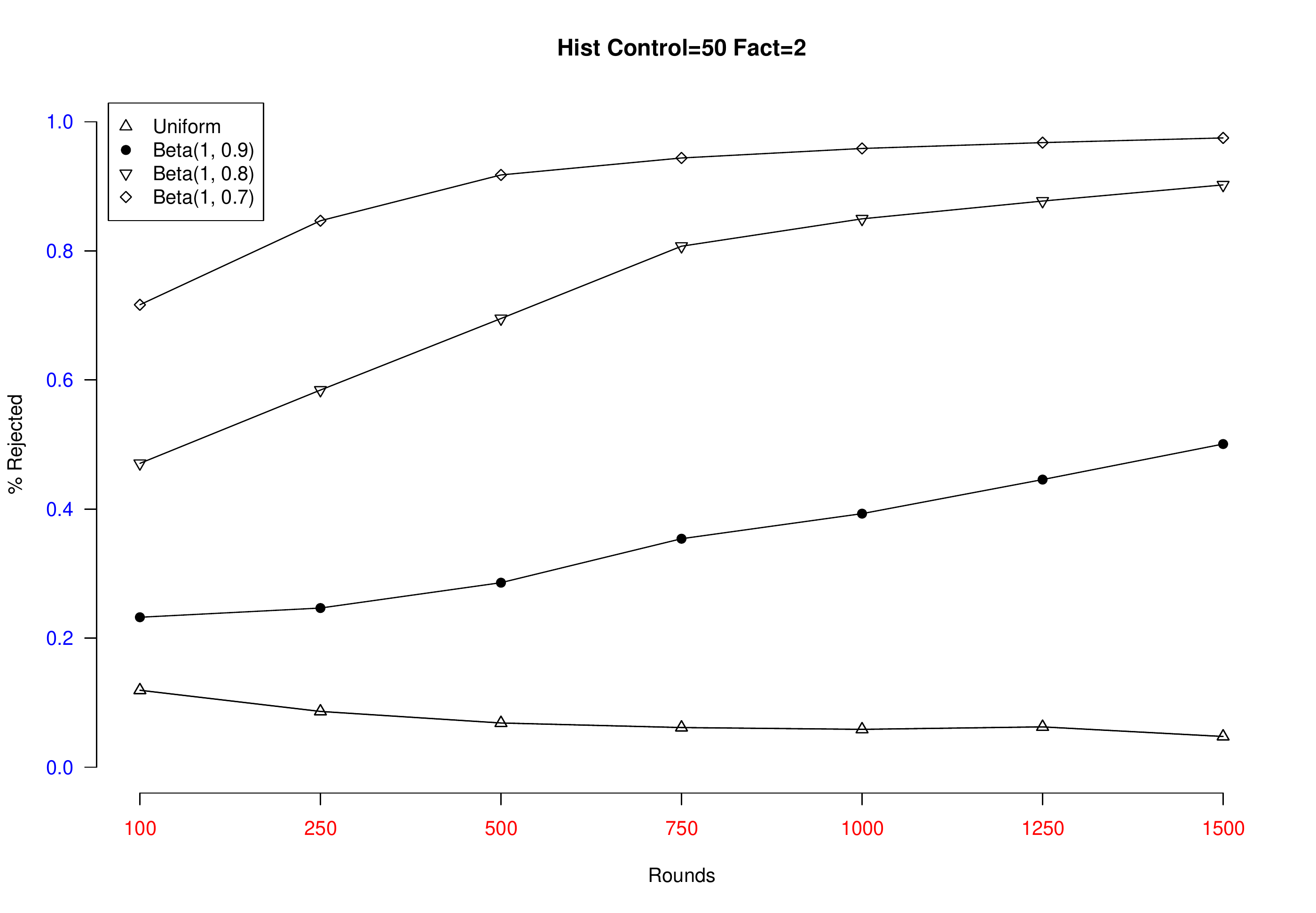}}
      \caption {Percentage of rejected values.}
      \label{fig:faultdetection}
\end{figure*}

Finally, for the same simulation scenario, in Fig.~\ref {fig:faultdetection} we compare behavior of the implementable KS test of Algorithm~\ref{alg:ImplementableQuidProQuo} for fair users (uniform) playing against a node with several manipulative profiles (Beta distribution variants) as the number of rounds increase. As it can be observed, the honest player rapidly gets her values to pass the test, while the dishonest gets into trouble rapidly because her values are rejected, even with distributions very similar to the uniform.

\section{Conclusion and future work}
\label{s:conclusions}

Throughout this paper, we have presented QPQ, an algorithm for optimal allocation and execution of tasks in a distributed environment with selfish behavior. Unlike many of the preexisting works, this algorithm proposes a mechanism that does not use payment or prior information on the behavior of the players. We have demonstrated that the algorithm is tolerant to dishonest, non rational or rationally limited behaviors without punishing fair users and rewarding players proportionally to their degree of truthfulness.

The proposed algorithm may be adapted using reasonable approximations so that it can be implemented in real networks with affordable computational and communication complexity. For all these reasons, we claim that this algorithm opens new horizons for the creation of novel computing frameworks where users can openly and effectively cooperate to achieve a common goal, based on the collaborative execution of simple atomic independent tasks.

Despite this, authors consider necessary to carry out further research to make QPQ robust to more sophisticated selfishness scenarios. For example, it would be necessary to consider cases in which players are not independent, and associate in groups trying to break the system's fairness.

Another aspect that should be extended is related to the notion of task utility. We have assumed that all nodes have an interest in having all tasks done. However, in a real environment, it is possible that only a subset of tasks are relevant for a given node. Hence, further work should be developed to relax some of the  QPQ hypotheses, and deal with this aspect.

To conclude, another aspect that may be improved is investigating GoF tests other than the KS to analyze if they can provide advantages for real implementations of the algorithm (for instance, using just a small set of samples to implement the acceptation test).

\bibliographystyle{IEEEtran}
\bibliography{IEEEabrv,QuidProQuo}
\end{document}